\documentclass[acmsmall,nonacm=true]{acmart}

\acmJournal{PACMPL}
\acmVolume{1}
\acmNumber{CONF} 
\acmArticle{1}
\acmYear{2018}
\acmMonth{1}
\acmDOI{} 
\startPage{1}

\setcopyright{none}

\usepackage{booktabs}   
\usepackage{subcaption} 
\usepackage{listings}
\usepackage{amsmath,amsthm,upgreek}
\usepackage{unixode}
\usepackage{stmaryrd}
\usepackage{xspace}

\newcommand\ie{\emph{i.e.}\ }
\newcommand\eg{\emph{e.g.}\ }

\include{defs}
\usepackage{color}
\definecolor{keywordcolor}{rgb}{0.7, 0.1, 0.1}   
\definecolor{tacticcolor}{rgb}{0.0, 0.1, 0.6}    
\definecolor{commentcolor}{rgb}{0.4, 0.4, 0.4}   
\definecolor{symbolcolor}{rgb}{0.0, 0.1, 0.6}    
\definecolor{sortcolor}{rgb}{0.1, 0.5, 0.1}      
\definecolor{attributecolor}{rgb}{0.7, 0.1, 0.1} 

\begin{document}

\title{Sealing Pointer-Based Optimizations Behind Pure Functions}

{\author{Daniel Selsam}
\email{daselsam@microsoft.com}
\affiliation{%
  \institution{Microsoft Research}
  \streetaddress{One Microsoft Way}
  \city{Redmond}
  \state{WA}
  \postcode{98052}
}}

{\author{Simon Hudon}
\email{simon.hudon@gmail.com}
\affiliation{%
  \institution{Carnegie Mellon University\thanks{This paper describes work performed while author Simon Hudon was at Microsoft Research.}}
  \streetaddress{5000 Forbes Avenue}
  \city{Pittsburgh}
  \state{PA}
  \postcode{15213}
}}

{\author{Leonardo de Moura}
\email{leonardo@microsoft.org}
\affiliation{%
  \institution{Microsoft Research}
  \streetaddress{One Microsoft Way}
  \city{Redmond}
  \state{WA}
  \postcode{98052}
}}

\renewcommand{\shortauthors}{Selsam et al.}

\begin{abstract}
  Functional programming languages are particularly well-suited for building automated reasoning systems,
  since (among other reasons) a logical term is well modeled by an inductive type,
  traversing a term can be implemented generically as a higher-order combinator,
  and backtracking search is dramatically simplified by persistent datastructures.
  However, existing \emph{pure} functional programming languages all suffer a major limitation in these domains:
  traversing a term requires time proportional to the tree size of the term as opposed to its graph size.
  This limitation would be particularly devastating when building automation for interactive theorem provers
  such as Lean~\cite{de2015lean} and Coq~\cite{coqv891},
  for which the exponential blowup of term-tree sizes has proved to be both common and difficult to prevent.
  All that is needed to recover the optimal scaling is the ability to perform simple operations on the memory addresses of terms,
  and yet allowing these operations to be used freely would clearly violate the basic premise of referential transparency.
  We show how to use dependent types to seal the necessary pointer-address manipulations behind pure functional interfaces while requiring
  only a negligible amount of additional trust. We have implemented our approach for the upcoming version (v4)
  of Lean, and our approach could be adopted by other languages based on dependent type theory as well.

\end{abstract}

\ccsdesc[500]{Software and its engineering~General programming languages}
\ccsdesc[300]{Social and professional topics~History of programming languages}


\maketitle

\section{Introduction}\label{sec:intro}
Functional programming languages are particularly well-suited for building automated reasoning systems
for several reasons. First, a logical term is naturally represented by an inductive type,
whereas such terms are notoriously awkward to encode in object-oriented languages.
Second, traversing a term can be implemented generically as a higher-order combinator,
and so much boilerplate control-flow code can be avoided.
Third, backtracking search is dramatically simplified by the use of persistent datastructures,
since branches can be paused and resumed at will.
Indeed most interactive theorem provers are written in functional programming languages:
Isabelle/HOL~\cite{nipkow2002isabelle} is written in Poly/ML~\cite{matthews1985poly},
Coq is written in OCaml~\cite{leroy2018ocaml},
Agda~\cite{bove2009brief} and Idris~\cite{brady2013idris} are both written in Haskell~\cite{jones2003haskell},
and Lean~\cite{de2015lean} was written in C++~\cite{ellis1990annotated}
but is being rewritten in Lean itself.

Functional programming languages shine in this domain, yet
to the best of our knowledge the \emph{pure} fragments of existing functional programming languages
such as Haskell~\cite{jones2003haskell},
Gallina~\cite{huet1992gallina} (\ie the language of Coq), Idris~\cite{brady2013idris},
Agda~\cite{bove2009brief}, Miranda~\cite{turner1986overview}, PureScript~\cite{freeman2015purescript}
and Lean all suffer a critical limitation:
traversing a term requires time proportional to the tree size of the term as opposed to its graph size.
This limitation is particularly devastating in automated reasoning where the basic operations
can and do produce terms
whose tree representations are exponentially larger than their graph representations.
Even a single first-order unification can produce such explosion in principle, with the canonical example unifying
\( f(x_1, \dotsc, x_n) \) with \( f(g(x_2, x_2), \dotsc, g(x_{n+1}, x_{n+1})) \)~\cite{goubault1994implementing}.
The problem is exacerbated when writing automation for interactive theorem provers such as Lean and Coq
since terms are often the result of long chains of user-written meta-programs (\ie tactics).
In Lean's mathematics library, \textsf{mathlib}~\cite{DBLP:conf/cpp/X20},
despite conscious effort to avoid idioms known to cause this kind of explosion (\eg those pointed out by~\cite{garillot2011generic}),
there are still proofs that contain only 20,000 nodes when viewed as graphs but 2.5 billion nodes when viewed as trees.

All that is needed to traverse terms in time proportional to their graph sizes rather than their tree sizes
is the ability to perform simple operations on their memory addresses.
However, allowing unrestricted use of these operations would clearly violate the basic premise of referential transparency.
In this work, we show how to use dependent types to seal the necessary pointer-address manipulations
behind pure functional interfaces while requiring only a negligible amount of additional trust.
Our work is particularly relevant for building high-performance systems for automated reasoning,
but the pointer-based optimizations we consider are ubiquitous in real-world software projects and may provide
performance improvements in diverse domains.

We assume a dependently typed language that gets compiled to a low-level imperative IR, and
our approach is based on the following insights. First, if a
function is treated as an opaque primitive throughout the compilation process all
the way down to the IR, the body of the
function can then be replaced with a low-level imperative version.
Second, since the compiler and runtime
of a language are already trusted, it requires very little additional trust to
assume that simple properties that the runtime relies on do indeed
hold, for example that two live objects with the same memory address must be equal.
Third, by making use of these assumptions one can often formulate sufficient conditions for
the replacement code for a given function to be faithful to the original pure definition.
These conditions can then be encoded formally using
dependent types and required as additional arguments to the functions in question.
Then by design every full application of the functions can be safely replaced in the IR with their low-level imperative versions.

We stress that our accelerated implementations are more than just type-safe:
they are functionally indistinguishable from the pure reference implementations.
Thus any theorem one proves about one's pure functional implementations
holds for the accelerated version as well.
We have implemented our approach for the upcoming version (v4)
of Lean, and our approach could be adopted by other languages based on dependent type theory as well.
Complete versions of all examples in the paper are available in the supplementary material.

\section{Preliminaries}\label{sec:prelim}

For our present purposes, the distinguishing feature of dependently typed programming languages is that
proofs are first-class in the language.
In particular, a function can take a proof as an argument,
thereby ensuring that it can never be fully applied
unless the corresponding precondition is satisfied.
We illustrate with the classic example of returning the head of a non-empty list:
\begin{lstlisting}[numbers=left]
def List.head : ∀ (xs : List α) (pf : xs ≠ []), α
| [], (pf : [] ≠ []) => absurd rfl pf
| x::_, _ => x
\end{lstlisting}
In addition to the list \lstinline|(xs : List α)|,
the function \lstinline|List.head| takes an additional argument \lstinline|(pf : xs ≠ [])|
constituting a proof that the list \lstinline|xs| is not empty.
Note that the type \lstinline|xs ≠ []| \emph{depends on} the term \lstinline|xs|, hence the
name \emph{dependent types}.
The function body starts by jointly pattern-matching on \lstinline|xs| and \lstinline|pf|.
In the \lstinline|[]| branch (Line 2), the type of \lstinline{pf} becomes \lstinline{[] ≠ []},
which contradicts the reflexivity of equality \lstinline{rfl : [] = []}.
The \lstinline{absurd} function takes two contradictory facts as inputs and
lets us produce a term of any type we wish, in this case \lstinline{α}.
Finally, in the non-empty branch (Line 3), the function ignores the proof and returns the head of the list.

To simplify the presentation,
we replace almost all proofs in the paper with the symbol \lstinline{#}---no
matter how trivial the proofs may be---and relegate their details to the supplementary material.
Equality-substitution proofs are an exception, and we think it improves readability to include them.
We use the notation \lstinline{▸} as follows:
if \lstinline{(x y : α) (p : x = y) (h : r x)}, then \lstinline{p ▸ h} is a proof of \lstinline{r y}.
Note that if there were multiple occurrences of \lstinline{x} in the type of \lstinline{h},
the subset of occurrences to substitute would be inferred from the context.

Our presentation makes use of the \emph{squash} type former
(also known as \eg the proposition truncation and the (-1)-truncation)
that turns any type into a subsingleton type, \ie a type with at most one element~\cite{hottbook}.
More precisely, for any type \lstinline{α} we can form the type
\lstinline{‖α‖} such that for any \lstinline{(x : α)}, \lstinline{|x|} has type \lstinline{‖α‖},
and \lstinline{∀ (x y : α), |x| = |y|}. If \lstinline{(β : Type)} is a subsingleton, then we
can \emph{lift} a function \lstinline{f : α → β} to a function \lstinline{Squash.lift f : ‖α‖ → β}
such that \lstinline{∀ (x : α), Squash.lift f |x| = f x}.
Squashing can be defined in terms of \emph{quotient types}
(see \eg~\citet{altenkirch2016type,Nog02b,Cohen13quotient,Isabelle2010quotient,hofmann1995extensional,hottbook}),
as the special case of quotienting by the trivial relation that always returns true.

In several places, we use the standard \lstinline{Unit} type with one trivial constructor:
\begin{lstlisting}
inductive Unit : Type
| () : Unit
\end{lstlisting}
Note that in Haskell, the \lstinline{()} notation is used for both the type \lstinline{Unit} and the value \lstinline{() : Unit}.
Our presentation is simplified by the use of the \emph{state monad}~\cite{wadler1990comprehending}
as is common in Haskell to weave (functional) state through computations conveniently:
\begin{lstlisting}
def StateM σ α := σ → α × σ

def get : StateM σ σ := λ s => (s, s)
def set (s : σ) : StateM σ Unit := λ _ => ((), s)
def modify (f : σ → σ) : StateM σ Unit := λ s => ((), f s)
def pure (x : α) : StateM σ α := λ s => (x, s)

def bind (c₁ : StateM σ α) (c₂ : α → StateM σ β) : StateM σ β :=
λ s => let (x, s) := c₁ s; c₂ x s

def modifySquash (f : α → α) : StateM ‖α‖ Unit :=
modify (Squash.lift (λ x => |f x|))
\end{lstlisting}
We also adopt Haskell's \emph{do} notation, so that \eg
\lstinline|do s ← get; set (f s); pure true|
is sugar for \lstinline|bind get (λ s => bind (set (f s)) (λ _ => pure true))|,
which itself is equivalent to \lstinline|λ s => (true, f s)|.

Our presentation is also simplified by the use of typeclasses~\cite{wadler1989make},
which are structures that can be synthesized automatically
by backward chaining~\cite{sozeau2008first,selsam2020tabled}.
A simple example is the class of types possessing a default element:
\begin{lstlisting}
class HasDefault (α : Type) : Type := (default : α)
\end{lstlisting}
with example instances:
\begin{lstlisting}
instance : HasDefault Nat := { default := 0 }
instance : HasDefault (Option α) := { default := none }
\end{lstlisting}
We can define a function \lstinline{default} that
takes a \lstinline{HasDefault α} instance as an \emph{instance-implicit} argument, indicated by
square brackets:
\begin{lstlisting}
def default (α : Type) [HasDefault α] : α := HasDefault.default α
\end{lstlisting}
Instance-implicit arguments do not need to be passed explicitly, and are instead
synthesized automatically by typeclass resolution based on the instances that have been registered.
For example, \lstinline{default Nat} will return \lstinline|0|,
whereas \lstinline|default (Option String)| will return \lstinline{none}.
The class of subsingletons is particularly useful in our setting:
\begin{lstlisting}
class Subsingleton (α : Type) : Prop := (h : ∀ (x y : α), x = y)
\end{lstlisting}
Recall from above that \lstinline{‖α‖} is a subsingleton for all types \lstinline{α}.
Another subsingleton that we use is the result of applying a function to a given input argument:
\begin{lstlisting}
structure Result (f : α → β) (x : α) : Type := (output : β) (h : output = f x)
\end{lstlisting}
We also use the fact that products of subsingletons are subsingletons,
and that functions mapping to subsingletons are themselves subsingletons.
Together these imply that if \lstinline{α} and \lstinline{β} are both subsingletons,
then so is the state monad computation \lstinline{StateM α β := α → β × α}.

We also need a type to represent decidable propositions:
\begin{lstlisting}
inductive Decidable (p : Prop) : Type
| isTrue  (h : p) : Decidable
| isFalse (h : ¬p) : Decidable
\end{lstlisting}
Note that since the parameter \lstinline{p} is necessarily a parameter of the types returned by the constructors,
it is not necessary to make this dependence explicit and we write \lstinline{Decidable} rather than \lstinline{Decidable p}.
Equality in dependent type theory is not in general decidable; whereas \lstinline|Bool| is the standard
two-element datatype from traditional programming languages, \lstinline|Prop| is the type of all propositions,
and not every proposition has a proof or a disproof (\eg by the Halting Problem). The \lstinline{Decidable} typeclass
lets us blur the distinction between \lstinline|Prop| and \lstinline|Bool| in the common case by projecting
decidable propositions to booleans automatically. We can make this conversion explicit with the function
\lstinline{toBool : Decidable p → Bool}, which satisfies the following basic properties:
\begin{lstlisting}
def toBool : Decidable p → Bool
| isTrue  _ => true
| isFalse _ => false

theorem toBoolEqTrue (d : Decidable p) (h : p) : toBool d = true
theorem ofToBoolEqTrue (d : Decidable p) (h : toBool d = true) : p
theorem ofToBoolEqFalse (d : Decidable p) (h : toBool d = false) : ¬ p
\end{lstlisting}
Note that different values of type \lstinline{Decidable p} may correspond to radically different algorithms
for deciding \lstinline{p}.
Although \lstinline{Decidable} is a typeclass in Lean, for our presentation it is more convenient
to always pass the \lstinline{Decidable} arguments explicitly.

Lastly, we need the following helper function for branching on a boolean with
access to equality proofs in both branches:
\begin{lstlisting}
def condEq (b : Bool) (h₁ : b = true → β) (h₂ : b = false → β) : β
\end{lstlisting}

\section{Pointer Equality Optimizations}\label{sec:equality}
\subsection{withPtrEq}\label{sec:equality:withPtrEq}
Imperative programmers routinely use pointer equality to accelerate reflexive binary relations such as structural equality.
Suppose we are evaluating a reflexive binary relation \lstinline{r : α → α → Bool} on two terms \lstinline{(t₁ t₂ : α)}.
If \lstinline{t₁} and \lstinline{t₂} have the same address in memory, then they must be the same object,
and hence \lstinline{r t₁ t₂} can safely return true without proceeding further.
However, this optimization is unsound if \lstinline{r} is not actually reflexive, and confirming that an arbitrary relation is reflexive
falls well beyond the capabilities of existing functional programming languages based on simple type theory.
Fortunately, languages based on dependent type theory can establish such properties at compile time, and so confirm that
particular uses of this trick are sound.

To support this idiom and others, we introduce the following new primitive:
\begin{lstlisting}
def withPtrEq (x y : α) (k : Unit → Bool) (h : x = y → k () = true) : Bool := k ()
\end{lstlisting}
Viewed as a pure function, it simply evaluates the thunk \lstinline{k} and returns the result.
We refer to this pure implementation as the function's \emph{reference implementation},
and our goal will be to replace the reference implementation in the low-level IR with a faster but still
functionally equivalent implementation.
The dependently-typed argument \lstinline{(h : x = y → k () = true)} represents a proof
that the thunk \lstinline{k} will return \lstinline{true} whenever \lstinline{x = y}.
Thus if \lstinline{withPtrEq} is ever fully applied,
and if we could somehow determine that its first two arguments were equal (\eg by pointer equality),
we could evaluate the thunk \lstinline{k} correctly by simply returning \lstinline{true}.

The pure reference implementation notwithstanding,
the compiler can treat this definition as a new opaque primitive until reaching the low-level imperative IR,
which has support for accessing the memory addresses of objects,
and which already relies on the assumption that two live objects with the same memory address must be equal.
Thus by chaining together this implicit assumption about the runtime
with the proof \lstinline{(h : x = y → k () = true)} provided
as argument to \lstinline{withPtrEq}, a simple meta-logical argument establishes the soundness
of replacing the opaque \lstinline{withPtrEq} in the IR with a version that immediately returns \lstinline{true} if the addresses of \lstinline{x} and \lstinline{y} are equal,
and evaluates the thunk if they are not.
The Lean compiler ensures auxiliary closures are not allocated at runtime for the parameter \lstinline{k}, and erases the proof \lstinline{h}.
More specifically, \lstinline{withPtrEq x y (λ _ => f x y) h} will be compiled into the following low-level IR code (presented as pseudocode):
\begin{lstlisting}
if ptrAddr x == ptrAddr y then true else f x y
\end{lstlisting}
The low-level IR is compiled to C in a straightforward manner,
and the supplementary material shows how to inspect the exact C code generated for all examples in the paper.

The \lstinline{withPtrEq} primitive can be used to accelerate the test of a reflexive binary relation. We can define
a function \lstinline{withPtrRel} in terms of \lstinline{withPtrEq} that takes a binary relation \lstinline{r}
along with a proof that the relation is reflexive,
and returns a pointer-equality-accelerated version whose reference implementation is identical to the
reference implementation of the original relation:
\begin{lstlisting}
def withPtrRel (r : α → α → Bool) (h : ∀ (x : α), r x x = true) : α → α → Bool :=
λ (x y : α) => withPtrEq x y (λ _ => r x y) (λ (p : x = y) => p ▸ h x)
\end{lstlisting}

\subsection{One-off pointer equality tests}\label{sec:equality:oneoff}
\begin{figure}
  \includegraphics[height=100pt]{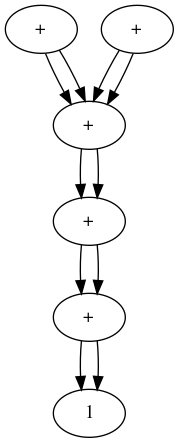}
  \caption{Two towers that are each of the form \lstinline{add (tower n) (tower n)},
    where all four towers are pointer equal but where the two outermost \lstinline{add} operations are not.
    On this example, the simplistic \lstinline{termEqOneOff} will only consider pointer equality at the
    respective roots, and so will take exponential despite the near-total sharing between the respective terms.
    We show in \S\ref{sec:equality:recursive} how to degrade gracefully in the presence of non-pointer-identical constructors
    by using \lstinline{withPtrEq} at each recursive call.
  }
  \label{fig:twotowersshared}
\end{figure}

Even a single application of \lstinline{withPtrEq} can provide exponential speedups in certain situations.
Consider the following simple term language:
\begin{lstlisting}
inductive Term : Type
| one  : Term
| add  : Term → Term → Term
\end{lstlisting}
along with the following function to generate a term tower:
\begin{lstlisting}
def tower : Nat → Term
| 0   => one
| n+1 => let t := tower n; add t t
\end{lstlisting}
\begin{figure}
\centering
\begin{subfigure}[b]{0.3\textwidth}
  \centering
    \includegraphics[height=100pt]{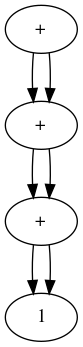}
    \caption{Graph}
    \label{fig:graph}
  \end{subfigure}
\begin{subfigure}[b]{0.6\textwidth}
    \centering
    \includegraphics[height=100pt]{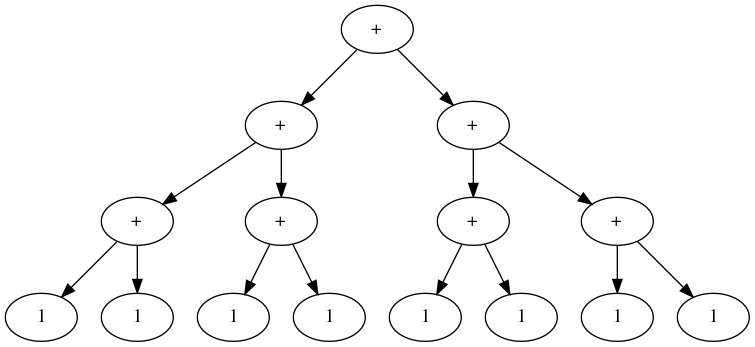}
    \caption{Tree}
    \label{fig:tree}
  \end{subfigure}
  \caption{Comparing the graph and tree representations of the term \lstinline{tower 4}.
    In general, \lstinline{tower n} has size $\Theta(n)$ as a graph but size $\Theta(2^n)$ as a tree.
    There is no way to traverse this term in sub-exponential time using the pure fragments of any existing languages.
    By sealing low-level pointer operations behind functional reference implementations,
    we recover the optimal $\Theta(n)$ scaling while preserving purity.}
  \label{fig:tower}
\end{figure}
Figure~\ref{fig:tower} shows both the graph and the tree representations of \lstinline{tower 4}.
The relevant point is that the size of the graph is $\Theta(n)$ whereas the size of the unfolded tree is $\Theta(2^n)$.
One of the main motivations of the present work is that
\emph{there is no way to traverse this term in sub-exponential time using existing pure functional languages}.
There are ways to construct other kinds of entities from the bottom up that are isomorphic to this term in the presence of some additional state
and that can be traversed efficiently, for example by either of the first two approaches described in~\cite{braibant2014implementing}.
However, in existing pure languages, there is no way to efficiently traverse a term of a standard inductive type like the one above.
For example, the following pure functional equality test will require $\Theta(2^n)$ time to even confirm that two pointer-identical
towers of height \lstinline{n} are equal:
\begin{lstlisting}
def termEqPure : Term → Term → Bool
| one, one => true
| add x₁ y₁, add x₂ y₂ => termEqPure x₁ x₂ && termEqPure y₁ y₂
| _, _ => false
\end{lstlisting}
Thus a single pointer equality test at the outset can provide exponential speedups on this problem
(once its reflexivity has been established):
\begin{lstlisting}
theorem termEqPureRefl : ∀ (t : Term), termEqPure t t = true

def termEqOneOff : Term → Term → Bool := withPtrRel termEqPure termEqPureRefl
\end{lstlisting}
However, any deviation from perfect sharing would cause the speedups from \lstinline{termEqOneOff} to evaporate.
Figure~\ref{fig:twotowersshared} shows two towers that are each of the form \lstinline{add (tower n) (tower n)},
where all four towers are pointer equal but where the two outermost \lstinline{add} operations are not.
Then since \lstinline{termEqOneOff t₁ t₂} falls back on \lstinline{termEqPure t₁ t₂} once its two arguments
are found not to be pointer equal, it will take exponential time to evaluate.
In order to degrade gracefully in the presence of non-pointer-identical constructors,
pointer equality must be checked at each recursive call.

\subsection{Recursive pointer equality tests}\label{sec:equality:recursive}

The primitive \lstinline{withPtrEq} introduced above is sufficient to support recursive pointer equality tests as well,
though the construction is more involved. We now show how to construct a recursively-accelerated equality test for
the simple term language of \S\ref{sec:equality:oneoff}.
The main complication is that in order for the outer call to establish that it is reflexive,
the recursive calls must return proofs that their results are reflexive.
We could accomplish this by requiring a thunk that returns a subtype \lstinline|{ b : Bool // x = y → b = true }|,
but since we are primarily interested in speeding up equality rather than an arbitrary reflexive relation,
we accomplish the same goal using the standard \lstinline{Decidable} type
introduced in \S\ref{sec:prelim}:
\begin{lstlisting}
inductive Decidable (p : Prop) : Type
| isTrue  (h : p) : Decidable
| isFalse (h : ¬p) : Decidable
\end{lstlisting}
First, we need the following generic helper function \lstinline{withPtrDecEq}
that tries to decide \lstinline{x = y} by passing a provided thunk to
\lstinline{withPtrEq}:
\begin{lstlisting}[numbers=left]
def withPtrEqDecEq (x y : α) (k : Unit → Decidable (x = y)) : Decidable (x = y) :=
let kb : Unit → Bool := λ _ => toBool (k ());
let kbRfl : x = y → kb () = true := toBoolEqTrue (k ());
let b : Bool := withPtrEq x y kb kbRfl;
condEq b
  (λ (h : b = true) => isTrue (ofToBoolEqTrue (k ()) h))
  (λ (h : b = false) => isFalse (ofToBoolEqFalse (k ()) h))
\end{lstlisting}
Whereas \lstinline{withPtrEq} takes a thunk returning \lstinline{Bool},
\lstinline{withPtrEqDecEq} takes a thunk \lstinline{k} returning a term of type \lstinline{Decidable (x = y)},
which constitutes both a boolean value (whether \lstinline{x} and \lstinline{y} are equal)
along with a proof that the boolean value is consistent with whether or not \lstinline{x} and \lstinline{y}
are actually equal.
First, \lstinline{withPtrEqDecEq} creates a boolean thunk \lstinline{kb} that can be passed to
\lstinline{withPtrEq}, that uses \lstinline{toBool} to extract the boolean
out of the \lstinline{Decidable (x = y)} value returned by the thunk \lstinline{k} (Line 2).
It then establishes the proof obligation for \lstinline{withPtrEq} using \lstinline{toBoolEqTrue} (Line 3)
and calls \lstinline{withPtrEq} (Line 4). Finally, \lstinline{condEq} is used to branch on the value of \lstinline{b} (Line 5),
and in each branch the proofs are lifted to terms of type \lstinline{Decidable (x = y)}
using basic lemmas (Lines 6-7). Note that we can pass \eg \lstinline{h : b = true} to \lstinline{ofToBoolEqTrue (k ())}
because the reference implementation of \lstinline{withPtrEq} simply evaluates the thunk \lstinline{k},
and so the result \lstinline{b} returned by \lstinline{withPtrEq} is definitionally equal to \lstinline{toBool (k ())}.

Next, we can define the continuation \lstinline{k} for decidable equality on \lstinline{Term}s:
\begin{lstlisting}
def termDecEqAux : ∀ (t₁ t₂ : Term), Decidable (t₁ = t₂)
| one,       one       => isTrue rfl
| add x₁ y₁, add x₂ y₂ =>
  match withPtrEqDecEq x₁ x₂ (λ _ => termDecEqAux x₁ x₂) with
  | isTrue h₁ =>
    match withPtrEqDecEq y₁ y₂ (λ _ => termDecEqAux y₁ y₂) with
    | isTrue h₂  => isTrue (h₁ ▸ h₂ ▸ rfl)
    | isFalse h₂ => isFalse #
  | isFalse h₁ => isFalse #
| one,       add x y   => isFalse #
| add x y,   one       => isFalse #
\end{lstlisting}
This version is almost identical to the naive version \lstinline{termEqPure},
except it calls \lstinline{withPtrEqDecEq} for all recursive calls (passing itself as the continuation),
and it also produces proofs in each of the branches that it is truly computing equality.
Finally, we wrap this auxiliary function with a top-level pointer equality check:
\begin{lstlisting}
def termDecEq : ∀ (t₁ t₂ : Term), Decidable (t₁ = t₂) :=
λ t₁ t₂ => withPtrEqDecEq t₁ t₂ (λ _ => termDecEqAux t₁ t₂)
\end{lstlisting}
and extract the Boolean equality test from it:
\begin{lstlisting}
def termEqRec (t₁ t₂ : Term) : Bool := toBool (termDecEq t₁ t₂)
\end{lstlisting}
This construction is only a minor variation of the
automatically-generated definitions already produced by pure functional languages
(\eg by Haskell's \lstinline{deriving (Eq)}).
Whereas \lstinline{termEqOneOff} only provides speedups when comparing pointer-equal towers,
\lstinline{termEqRec} provides speedups exponential in the height of the shared pointer-equal base of two structurally equal towers,
no matter how many non-pointer-equal constructors wrap the respective bases.
Although this is an improvement over \lstinline{termEqOneOff}, it will still take exponential time to determine that two pointer-disjoint towers
of the same height are structurally equal. We revisit this scenario in \S\ref{sec:traverse:arbitrary}.

\section{Traversing Terms in Linear Time}\label{sec:traverse}

We now show how to use the pointer equality optimizations discussed in \S\ref{sec:equality}
to traverse terms in linear time. As a running example, we consider the function that evaluates a \lstinline{Term} into a natural number:
\begin{lstlisting}
def evalNatNaive : Term → Nat
| one => 1
| add t₁ t₂ => evalNatNaive t₁ + evalNatNaive t₂
\end{lstlisting}
This section improves on the na\"{i}ve version incrementally and culminates in \S\ref{sec:traverse:arbitrary} with the implementation of \lstinline{evalNatRobust},
which scales linearly in the graph size rather than the tree size no matter the memory layout of the term.

\subsection{Pure functional hash maps}\label{sec:traverse:purehash}

Pure functional hash maps---also called hash trees, hash tries, persistent hash maps, and hash array mapped tries---are
a common datastructure in functional programming languages.
They were introduced by \citet{bagwell2001ideal}
and are now part of the standard library in Lean4, Clojure~\cite{hickey2008clojure} and Scala~\cite{odersky2004overview}.
They are also included in the \lstinline{unordered-containers} package in Haskell.
Finding, inserting and deleting each technically require $O(\log_{B}(n))$ time for a branching factor $B = 2^k$,
though \citet{bagwell2001ideal} simplifies this to $O(1)$ in his analysis.

Many functional languages based on reference counting---including Lean4,
PVS~\cite{owre1992pvs}, SISAL~\cite{sisal}, and SAC~\cite{sac}---also
support traditional hash maps that have the desired (amortized) $O(1)$ cost per operation
as long as the map is not shared, \ie its reference count is 1.
In particular,
the Lean4 standard library includes a hash map based on an array of buckets,
and thanks to the optimizations described in \citet{ullrich2019counting}, the array will be updated destructively
as long as the hash map is used linearly, which it is in all the examples that follow.
For languages that do not support such destructive updates, the approach we now describe will allow
traversing terms in either linear time or quasilinear time, depending on whether or not $O(\log_B(n))$ is considered $O(1)$.

\subsection{Intrusive hash codes}\label{sec:traverse:intrusive}

A naive implementation of hashing a term requires a traversal
and hence a single call will take exponential time on \lstinline{tower n}.
However, since hashing is a (pure) \emph{unary} function of a term,
we can hash terms in constant time by simply extending the \lstinline{Term} type to store its hash code:
\begin{lstlisting}
inductive Term : Type
| one  : Term
| add  : Term → Term → Addr → Term

def fastHash : Term → Addr
| one => 7
| add t₁ t₂ h => h

def add (t₁ t₂ : Term) : Term :=
Term.add t₁ t₂ (mixHash (fastHash t₁) (fastHash t₂))
\end{lstlisting}
where \lstinline{Addr} is a fixed-size numeric type that is big enough to store any pointer address.
Alternatively, if there were more constructors, it may be more convenient to define a new type that packages a \lstinline{Term},
a hash code, and (optionally) a proof that the stored hash code indeed agrees with the naive hash of the term:
\begin{lstlisting}
inductive TermNoHash : Type
| one  : TermNoHash
| add  : TermNoHash → TermNoHash → TermNoHash

def slowHash : TermNoHash → Addr

structure Term : Type :=
(term : TermNoHash) (hash : Addr) (hashOk : hash = slowHash term)

def fastHash : Term → Addr
| Term term hash hashOk => hash

def add : Term → Term → Term
| Term t₁ h₁ ok₁, Term t₂ h₂ ok₂ =>
  Term (TermNoHash.add t₁ t₂) (mixHash (fastHash t₁) (fastHash t₂)) #
\end{lstlisting}

We advocate intrusive hashing for most use cases, though
we present an alternative that does not rely on it in \S\ref{sec:extensions}.
For the rest of \S\ref{sec:traverse}, \lstinline{Term} will refer to the first variant above, with its
\lstinline{add} constructor intrusively extended to include its hash code.
Moreover, we assume that this field is always compared before the children inside \lstinline{termEqRec}.

\subsection{Traversing near-perfect towers}\label{sec:traverse:perfect}

\begin{figure}
  \includegraphics[height=100pt]{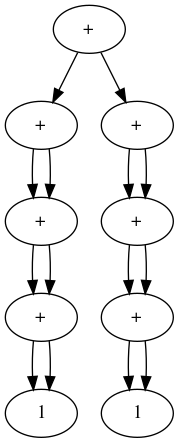}
  \caption{A term of the form \lstinline{add (tower n) (tower n)} where the two towers are pointer-disjoint.
    The simple caching approach of \S\ref{sec:traverse:perfect} will take exponential time on this example.
    \S\ref{sec:traverse:arbitrary} presents the general solution that scales in the graph size rather than the term size no matter the shape of the term.}
  \label{fig:twotowersdisjoint}
\end{figure}

By caching with a hash table that combines \lstinline{termEqRec} with the intrusive hash code (\S\ref{sec:traverse:intrusive}),
we can evaluate functions on both the tower of Figure~\ref{fig:graph}
and the near-perfect tower of Figure~\ref{fig:twotowersshared} in (expected) linear time.
For example, the following function that evaluates a \lstinline{Term} as a natural number runs in linear
time:\footnote{Note that while Coq's trusted termination checker would accept the analogous Coq program as written,
  Lean requires well-founded recursion for this example and in particular generates proof obligations for the recursive calls.
  However, the example could be expressed (if less clearly) in terms of direct structural recursion only.}
\begin{lstlisting}
def evalNat : Term → StateM (HashMap Term Nat) Nat
| t => do
  map ← get;
  match map.find? t with
  | some n => pure n
  | none =>
    match t with
    | one => pure 1
    | add t₁ t₂ hash => do
      n₁ ← evalNat t₁;
      n₂ ← evalNat t₂;
      let n := n₁ + n₂;
      modify (λ map => map.insert t n);
      pure n
\end{lstlisting}
It is not important that the function returns a scalar.
On the two example terms above, this approach will scale with the graph size rather than the term size
even if the function returns a new term, and even if the function recurses on (shallow) combinations of existing subterms---
for example, if we add a \lstinline{mul} constructor and distribute multiplication over addition
with \eg \lstinline{distrib (mul t (add t₁ t₂))} reducing to \lstinline{add (distrib (mul t t₁)) (distrib (mul t t₂))}).
However, this approach will still take exponential time when traversing a term that contains two pointer-disjoint towers,
like the term in Figure~\ref{fig:twotowersdisjoint}. Specifically, \lstinline{evalNat} will evaluate the first tower efficiently,
but then simply looking up the root of the second tower in the cache will fall back on structural equality, which in the absense of
any pointer equalities will take exponential time.
This limitation is similar to one alluded to at the end of \S\ref{sec:equality:recursive}.
\S\ref{sec:traverse:arbitrary} presents the general solution that scales in the graph size rather than the term size no matter the shape of the term.

\subsection{Traversing arbitrary terms}\label{sec:traverse:arbitrary}

We saw in \S\ref{sec:traverse:perfect} that as long as a term is (nearly) maximally shared,
we can traverse it in linear time by caching with a hash table that uses pointer-accelerated equality
and the intrusive hash. Thus, to traverse arbitrary terms in linear time, it suffices to be able
to make a term maximally shared in linear time. We refer to this process as \emph{sharing the common data} within a term,
and it is akin to hash-consing after-the-fact. Terms are traversed and normalized in a bottom-up fashion while building
a map from addresses of terms to addresses of equivalent but maximally-shared terms.
Although the additional primitives we will introduce in \S\ref{sec:extensions} allow implementing such a
sharing function that scales linearly in the graph size of the terms,
many runtimes (including Poly/ML\footnote{\url{https://www.polyml.org/documentation/Reference/PolyMLStructure.html\#shareCommonData}. Accessed 5/30/2020.} and Lean) already include a generic, high-performance implementation of it
that applies to terms of any type.
Thus we can apply the same approach we took in \S\ref{sec:equality:withPtrEq}
to seal the low-level implementation for sharing common data behind the (pure) polymorphic identity function.
Specifically, we introduce a new primitive
\begin{lstlisting}
def shareCommon (x : α) : α := x
\end{lstlisting}
Viewed as a pure function, it is simply the identity function, yet just as for \lstinline{withPtrEq},
the compiler treats this definition as a new opaque primitive until reaching the low-level imperative IR,
at which point it replaces it with a call to the runtime's \lstinline{shareCommon} function.
Since the correctness of the system already depends on the runtime's \lstinline{shareCommon} implementation
being functionally equivalent to the identity function,
this transformation only requires a negligible amount of additional trust.
Note that although this primitive does not require dependent types to be sealed by a pure function,
it would not affect the asymptotics of traversing terms on its own without the additional ability to
compare memory addresses.

By preceding the caching traversal of \S\ref{sec:traverse:perfect} with a call to \lstinline{shareCommon},
we can traverse an arbitrary term \lstinline{t : Term} in linear time. For example:
\begin{lstlisting}
def evalNatRobust (t : Term) : Nat := (evalNat (shareCommon t) HashMap.empty).1
\end{lstlisting}
In practice, it is wasteful to apply \lstinline{shareCommon} from scratch each time, since
the map from addresses of terms to addresses of equivalent but maximally-shared terms
will need to be rebuilt from scratch, and
each term will need to be traversed again even if it was already made maximally
shared by a previous \lstinline{shareCommon} call.
To accommodate incrementally sharing the common data across multiple terms,
we introduce a new primitive type \lstinline{ShareCommon.State : Type} (that wraps the map described above)
and the more general \lstinline{withShareCommon}:
\begin{lstlisting}
def ShareCommon.State : Type := Unit -- wraps the map from addresses to addresses
def ShareCommon.State.empty : ShareCommon.State := ()
def withShareCommon (x : α) : StateM ShareCommon.State α := pure x
\end{lstlisting}
Here \lstinline{withShareCommon} is a primitive that behaves like \lstinline{shareCommon} above
except it starts sharing the common data given the state it is passed and then returns the resulting state.
Using the new primitive \lstinline{withShareCommon}, we can now define the original \lstinline{shareCommon} as
\begin{lstlisting}
def shareCommon (x : α) : α := (withShareCommon x ShareCommon.State.empty).1
\end{lstlisting}

In Lean, \lstinline{withShareCommon} satisfies the desirable property that
\lstinline{do x ← withShareCommon x; x ← withShareCommon x; f x} is algorithmatically equivalent to \lstinline{do x ← withShareCommon x; f x},
since the second \lstinline{withShareCommon} call will necessarily be a no-op.
However, this property will not hold in general for languages such as OCaml and Haskell that use a moving (also known as compacting) garbage collector
since objects may be moved at any time.

\section{Extensions}\label{sec:extensions}

In \S\ref{sec:traverse}, we saw how to combine \lstinline{withPtrEq}, intrusive hash codes, and a \lstinline{shareCommon} primitive
to traverse arbitrary terms in (expected) linear time while preserving functional equivalence with respect to a pure reference implementation.
In our experience building automation for earlier versions of Lean, we found that unsafe versions of the methods in \S\ref{sec:traverse} yield very
good performance in all our uses cases.
Nonetheless, we now introduce two extensions to \lstinline{withPtrEq} that may provide desirable trade-offs in certain contexts:
\lstinline{withPtrEqResult} of \S\ref{sec:extensions:withPtrEqResult} allows giving up
rather than recursing in the absence of pointer equality, while
\lstinline{withPtrAddr} of \S\ref{sec:extensions:withPtrAddr} allows using memory addresses directly as hash codes.
We integrate these two extensions to implement \lstinline{evalNatPtrCache} in \S\ref{sec:extensions:traversepointer},
which is an alternative to \lstinline{evalNatRobust} (\S\ref{sec:traverse:arbitrary}) that can traverse terms in linear time without
requiring intrusive hashes nor a call to the \lstinline{shareCommon} primitive.
As we will see, a notable downside of this alternative approach is that it requires a reference implementation for the function being cached.
For this reason among others, we generally advocate the approach of \S\ref{sec:traverse}.

\subsection{Imprecise equality tests}\label{sec:extensions:withPtrEqResult}

One limitation of the approach of \S\ref{sec:traverse} is that even when a programmer
knows that a term must be maximally shared in a particular context,
it will still recurse into subterms when pointer equality fails to hold
for two elements in the same hash bucket.
However, this is rarely an issue in practice since it is highly unlikely that the hash codes of the subterms will also collide,
and so \lstinline{termEqRec} will still fail quickly.
Nonetheless, we show how to apply the same methodology of \S\ref{sec:equality:withPtrEq} to seal an \emph{imprecise}
pointer equality test---one that gives up rather than
recurses in the absence of pointer equality---behind a pure functional interface.
Of course, arbitrary uses of imprecise pointer equality tests will not be sound in general.
However, a use is clearly sound if the continuation returns an element of a subsingleton type (see \S\ref{sec:prelim}),
since there is only one value it could possibly return no matter how it computes the value internally.
It turns out that this simple precondition is expressive enough to support our current needs.

To support imprecise equality tests, we define a new inductive type for the result of a pointer equality test:
\begin{lstlisting}
inductive PtrEqResult (x y : α) : Type
| unknown  : PtrEqResult
| yesEqual : x = y → PtrEqResult
\end{lstlisting}
and introduce a second primitive, \lstinline{withPtrEqResult}:
\begin{lstlisting}
def withPtrEqResult [Subsingleton β] (x y : α) (k : PtrEqResult x y → β) : β :=
k unknown
\end{lstlisting}
This primitive differs from the original \lstinline{withPtrEq} in two ways.
First, rather than returning a boolean, the continuation \lstinline{k} can instead
return a subsingleton type \lstinline{β}. Second, rather than taking an argument of type \lstinline{Unit},
the continuation either gets no information (\lstinline{unknown})
or a proof that the two elements are equal (\lstinline{yesEqual (h : x = y)}).
We will see shortly why this proof is necessary.
The reference implementation simply evaluates the continuation \lstinline{k} on \lstinline{unknown}.
As for \lstinline{withPtrEq}, the compiler can treat this definition as a new opaque primitive until reaching the low-level imperative IR.
At this point it can replace the implementation with code that first checks pointer equality, and
then calls the continuation \lstinline{k} on either \lstinline{unknown} or \lstinline{yesEqual} depending on the result.
More specifically, Lean will compile \lstinline{withPtrEqResult x y k} into the following low-level IR code (presented as pseudocode):
\begin{lstlisting}
if ptrAddr x == ptrAddr y then k yesEqual else k unknown
\end{lstlisting}
Note that the \lstinline{yesEqual} is just a constant for the runtime, as the proof itself has no runtime representation and is erased by the compiler.
The soundness argument is similar to the one for \lstinline{withPtrEq}.
The runtime already relies on the assumption that two live objects with the same memory address must be equal.
Thus, when pointer equality is detected and \lstinline{k} is evaluated on \lstinline{yesEqual}, \lstinline{x} really does equal \lstinline{y}.
Moreover, since \lstinline{k} returns a subsingleton, the same result (up to equality)
will be returned no matter whether pointer equality is detected or not.
Thus the low-level imperative version is functionally equivalent to the pure reference implementation.

\paragraph{Imprecise association list caches.}
We now show how to implement an imprecise association list cache for a function \lstinline{f} using \lstinline{withPtrEqResult}.
Recall the subsingleton \lstinline{Result} type from \S\ref{sec:prelim}:
\begin{lstlisting}
structure Result (f : α → β) (x : α) : Type := (output : β) (h : output = f x)
\end{lstlisting}
and define an entry of the association list to be a dependent pair of an input \lstinline{x} and a \lstinline{Result f x}:
\begin{lstlisting}
structure Entry (f : α → β) : Type := (input : α) (result : Result f input)
\end{lstlisting}
We implement a function that looks up a \lstinline{Result} in an association list of \lstinline{Entry}s as follows:
\begin{lstlisting}[numbers=left]
def evalReadImpreciseListCacheOneOff (x₀ : α) : List (Entry f) → Result f x₀
| [] => Result.mk (f x₀) rfl -- rfl is the reflexivity proof of type f x₀ = f x₀
| (Entry.mk x r)::es =>
  withPtrEqResult x x₀ (λ (pr : PtrEqResult (x = x₀)) =>
    match x, pr, r with
    | _, yesEqual rfl, r => r
    | _, unknown _, _ => evalReadImpreciseListCacheOneOff es)
\end{lstlisting}
The three-way \lstinline{match} at the end may seem strange at first to readers unfamiliar with dependent pattern matching; we will explain how it works shortly.
If the list is empty, we simply evaluate \lstinline{f x₀} and return the result (Line 2).
Otherwise (Line 3), we perform an imprecise pointer equality test on \lstinline{x₀}
and the input \lstinline{x} of the first entry (Line 4). The continuation then
simultaneously pattern matches on \lstinline{x},
the result of the pointer equality test \lstinline{pr : PtrEqResult (x = x₀)}
and the result \lstinline{r : Result f x} (Line 5).
In the first branch (Line 6), \lstinline{pr} matches its second constructor, \lstinline{yesEqual},
and the argument to \lstinline{yesEqual} which would normally have type \lstinline{x = x₀} is able to reduce
to the reflexivity proof \lstinline{rfl : x₀ = x₀} since we are matching on \lstinline{x} simultaneously.
In this branch, \lstinline{r} has type \lstinline{Result f x₀} and so it suffices to return it.
In the branch where \lstinline{pr} does not contain a proof (Line 7),
it simply recurses on the rest of the list.
Note that there are no proof obligations besides the subsingleton requirement which is discharged by typeclass resolution.

The implementation above of \lstinline{evalReadImpreciseListCacheOneOff} has two limitations.
First, it only reads the list and does not return a new list on a cache miss.
It cannot simply return the modified list in addition to the result,
since \lstinline{withPtrEqResult} requires that the return type be a subsingleton.
We can address this limitation by taking an additional argument \lstinline{g : List (Entry f) → γ}
for some subsingleton \lstinline{γ}, and returning \lstinline{g} applied to the extended list
in addition to the result.
Second, it directly applies the function \lstinline{f} on a cache miss, and cannot be made to query
the pointer cache recursively on subterms.
We can address this limitation by taking a continuation as an argument that itself may read and write to the cache.
We present this version using the state monad \lstinline{StateM} to simplify the notation (see \S\ref{sec:prelim}):
\begin{lstlisting}[numbers=left]
def evalImpreciseBucket [Subsingleton γ] (x₀ : α) (k : StateM γ (Result f x₀))
  (update : Entry f → StateM γ Unit) : List (Entry f) → StateM γ (Result f x₀)
| [] => do
  r ← k;
  update (Entry.mk x₀ r);
  pure r
| (Entry.mk x r)::es =>
  withPtrEqResult x x₀ (λ (pr : PtrEqResult (x = x₀)) =>
    match x, pr, r with
    | _, yesEqual rfl, r => pure r
    | _, unknown _, r => evalImpreciseBucket es)
\end{lstlisting}
This function is the same as \lstinline{evalReadImpreciseListCacheOneOff}
except for two small changes to address the two limitations discussed in the previous paragraph.
First, to allow querying the cache recursively on subterms,
it calls the user-provided \lstinline{k} to compute the result on a cache-miss (Line 4),
rather than computing the function \lstinline{f} from scratch. Second, to allow returning an updated association list,
it applies the user-provided \lstinline{update} function to the computed result, which may update the subsingleton
state \lstinline{γ} as appropriate. We will see an example use of \lstinline{evalImpreciseBucket} in \S\ref{sec:extensions:withPtrAddr}.

\subsection{Pointer address hashing}\label{sec:extensions:withPtrAddr}

The intrusive approach to hashing presented in \S\ref{sec:traverse:intrusive}
is simple and effective, and yet it may not be the best solution in all contexts.
First, depending on the size of objects and the specifics of the runtime,
the intrusive hash codes might impose an undesirable space overhead.
Second, the intrusion imposes additional bookkeeping, both when defining the type and when proving properties about the program.
Third, for some workloads it can be difficult to design a good structural hash function.
Finally, in some situations it may be necessary to efficiently traverse existing terms of a type
that lacks an intrusive hash, if only to convert these terms to a type that has one.

To support direct pointer address manipulations, we introduce the following new primitive:
\begin{lstlisting}
def withPtrAddr [Subsingleton β] (x : α) (k : Addr → β) : β := k 0
\end{lstlisting}
where \lstinline{Addr} is a fixed-size numeric type that is big enough to store any pointer address.
The reference implementation of \lstinline{withPtrAddr} simply calls the continuation \lstinline{k} on the null address \lstinline{0},
but as usual, the compiler can treat this definition as a new opaque primitive until reaching the low-level imperative IR,
at which point it can evaluate \lstinline{k} on the actual memory address of \lstinline{x} rather than the null address.
More specifically, Lean will compile \lstinline{withPtrAddr x k} into the following low-level IR (pseudo)code: \lstinline{k (ptrAddr x)}.
Since the return type \lstinline{β} is a subsingleton,
\lstinline{k} will return the same result no matter what address it is evaluated on.
Thus the low-level version is functionally equivalent to the reference implementation.

\paragraph{Pointer caches.}
We now show how to use \lstinline{withPtrAddr}
to implement a cache that uses pointer addresses as hash codes.
To simplify the presentation, we will implement a simple array-based hash map,
though the same approach could be used to implement a pure functional hash map as well.
Resizing is also straightforward and so we omit it from our presentation.
We will use \lstinline{evalImpreciseBucket} (\S\ref{sec:extensions:withPtrEqResult}) for searching within each bucket,
so that structural equality is avoided altogether.
We first define a pointer cache for a function to be a squashed array of lists of entries for that function:
\begin{lstlisting}
def PtrCache (f : α → β) : Type := ‖Array (List (Entry f))‖
\end{lstlisting}
Since we squash the array in the definition of \lstinline{PtrCache}, \lstinline{PtrCache f} is a subsingleton type for any \lstinline{f}.
When we query a \lstinline{PtrCache f}
for a given value \lstinline|(x : α)|, we will return an element of the subsingleton type \lstinline{Result f x × PtrCache f},
so that we may inspect pointer addresses freely using \lstinline{withPtrAddr}.
Note that we are able to squash the array in the definition of \lstinline{PtrCache} because we only query the \lstinline{PtrCache} for subsingletons,
and so the actual contents of the underlying array cannot affect the values we compute.

We implement the function as follows:
\begin{lstlisting}
def evalPtrCache (x : α) (k : StateM (PtrCache f) (Result f x))
  : StateM (PtrCache f) (Result f x) := do
s ← get;
withPtrAddr x (λ u =>
  Squash.lift s (λ buckets =>
    if buckets.size = 0 then k else do -- alt: store proof of nonempty in PtrCache type
      let i := u.toNat % buckets.size;
      let update (e : Entry f) : StateM (PtrCache f) Unit :=
        modifySquash (λ buckets => Array.modify buckets i (λ es => e :: es));
      let es := Array.get! buckets i;
      evalImpreciseBucket x k update es))
\end{lstlisting}
As in \S\ref{sec:extensions:withPtrEqResult},
all of the proof obligations are reduced to establishing various types are subsingletons,
and are discharged automatically by typeclass resolution.

Note that Lean uses reference counting, and so the address of an object is constant.
Thus if a particular value \lstinline{(x : α)}
is inserted into a pointer cache, it will always be found when queried
in the future. However, this invariant does not hold in languages with a moving (also known as compacting) garbage collector,
and so there is a risk that a particular value \lstinline{(x : α)} may be re-inserted into multiple different
buckets without ever being found.
Although this is only a performance risk and cannot affect referential transparency, it constitutes
an additional reason for preferring the approach of \S\ref{sec:traverse}.

\subsection{Traversing terms with pointer address hashing}\label{sec:extensions:traversepointer}

We now show how to use \lstinline{evalPtrCache} from \S\ref{sec:extensions:withPtrAddr}
to traverse a term in linear time without the intrusive hash.
We consider the running example of \S\ref{sec:traverse} of evaluating a term into a natural number:
\begin{lstlisting}
def evalNatNaive : Term → Nat
| one => 1
| add t₁ t₂ => evalNatNaive t₁ + evalNatNaive t₂
\end{lstlisting}
It is convenient to give a name to the pointer address caching monad for a function \lstinline{f}:
\begin{lstlisting}
def PtrCacheM (f : α → β) (x : α) := StateM (PtrCache f) (Result f x)
\end{lstlisting}
Now we can implement \lstinline{evalNatPtrCache} using the tools of \S\ref{sec:extensions:withPtrEqResult} and \S\ref{sec:extensions:withPtrAddr} as follows:
\begin{lstlisting}[numbers=left]
def evalNatPtrCache : ∀ (t : Term), PtrCacheM evalNatNaive t
| one => pure (Result.mk 1 rfl) -- `1 = evalNatNaive one` by definition
| add t₁ t₂ => do
  Result.mk r₁ hr₁ ← evalPtrCache t₁ (evalNatPtrCache t₁);
  Result.mk r₂ hr₂ ← evalPtrCache t₂ (evalNatPtrCache t₂);
  let output : Nat := r₁ + r₂;
  let h : output = evalNatNaive t₁ + evalNatNaive t₂ := hr₁ ▸ hr₂ ▸ rfl;
  pure (Result.mk output h)
\end{lstlisting}
If the term is \lstinline{one} (Line 2), then it returns the number \lstinline{1}
along with a proof that \lstinline{1 = evalNatNaive one}, which is \lstinline{rfl}
since it holds by definition.
Otherwise, if the term is \lstinline{add t₁ t₂} (Line 3),
it first searches for \lstinline{t₁} and \lstinline{t₂} in the pointer cache (Lines 4-5).
For each child, it passes itself applied to that child as the pointer cache continuation,
so if the child is not in the pointer cache, \lstinline{evalNatPtrCache} will be called recursively on the child.
Then it sums the resulting values together (Line 6),
proves that the result is indeed faithful to \lstinline{evalNatNaive} (Line 7),
and bundles the output and the proof to return an element of type \lstinline{Result evalNatNaive t} (Line 8).
We are making use of the fact that
\lstinline{evalNatNaive (add t₁ t₂) = evalNatNaive t₁ + evalNatNaive t₂} holds by definition;
this step would need to be stated and proved explicitly if using a more sophisticated reference implementation.

We note that \lstinline{evalNatPtrCache} has an interesting advantage over the \lstinline{evalNat} from \S\ref{sec:traverse:perfect}:
it will scale linearly on the example from Figure~\ref{fig:twotowersdisjoint} without needing to preceed it by \lstinline{shareCommon},
since it will effectively cache the two different towers separately.

\section{Experiments}\label{sec:experiments}

We evaluate eight different variants of \lstinline{evalNat} on two different towers.
The eight variants we consider are as follows:

\begin{enumerate}
\item \lstinline{evalNatNoCache}: traverses recursively with no caching.
\item \lstinline{evalNatCacheSlowEqSlowHash}: caches but with na\"{i}ve equality tests and na\"{i}ve (unintrusive) hashing.
\item \lstinline{evalNatCacheSlowEqFastHash}: caches with the intrusive hash of \S\ref{sec:traverse:intrusive} but na\"{i}ve equality tests.
\item \lstinline{evalNatCacheFastEqSlowHash}: caches with the recursively-accelerated equality \lstinline{termEqRec} of \S\ref{sec:equality:recursive}
  but na\"{i}ve unintrusive hashing.
\item \lstinline{evalNatCacheFastEqFastHash}: caches with the recursively-accelerated equality \lstinline{termEqRec} of \S\ref{sec:equality:recursive}
  and the intrusive hash of \S\ref{sec:traverse:intrusive}. This variant corresponds to the \lstinline{evalNat} implementation in \S\ref{sec:traverse:perfect}.
\item \lstinline{evalNatCacheFastEqFastHashRobust}: the robust version of \lstinline{evalNatCacheFastEqFastHash} that first shares the common data.
  This variant corresponds to \lstinline{evalNatRobust} in \S\ref{sec:traverse:arbitrary}.
\item \lstinline{evalNatPtrCache}: uses the pointer cache described in \S\ref{sec:extensions:traversepointer}, and corresponds to \lstinline{evalNatPtrCache} of the same section.
\item \lstinline{evalNatPtrCacheRobust}: the same as \lstinline{evalNatPtrCache} after first sharing the common data.
\end{enumerate}

\begin{figure}
  \includegraphics[width=0.7\textwidth]{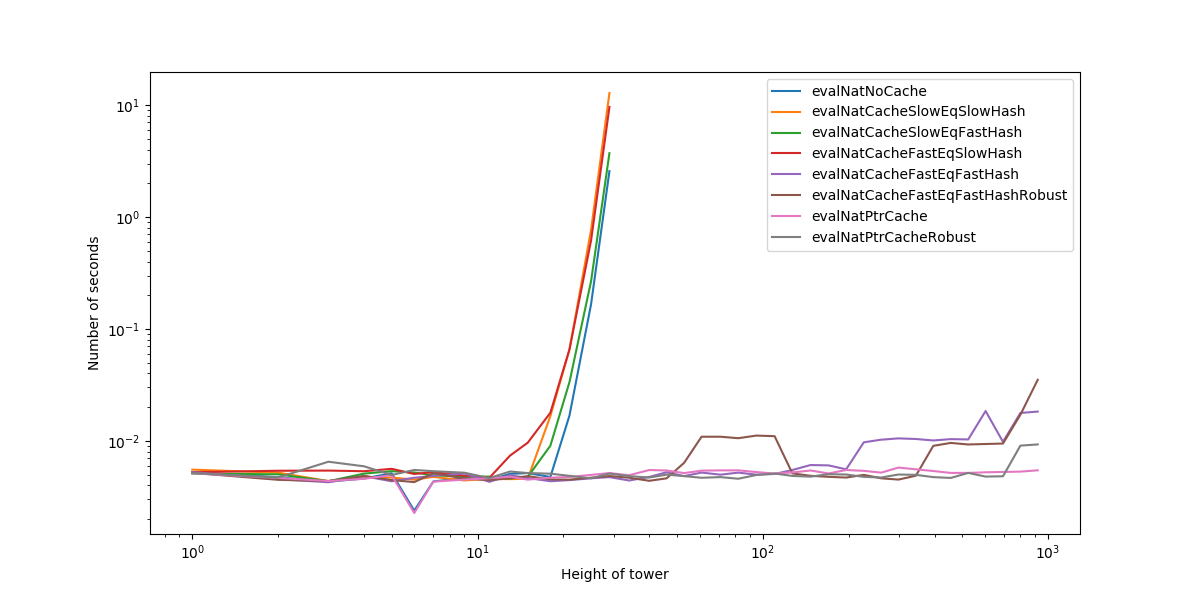}
  \caption{Comparing eight different variants of \lstinline{evalNat} on the simple maximally-shared tower of Figure~\ref{fig:tower}.}
  \label{fig:experiments:tower}
\end{figure}

\begin{figure}
  \includegraphics[width=0.7\textwidth]{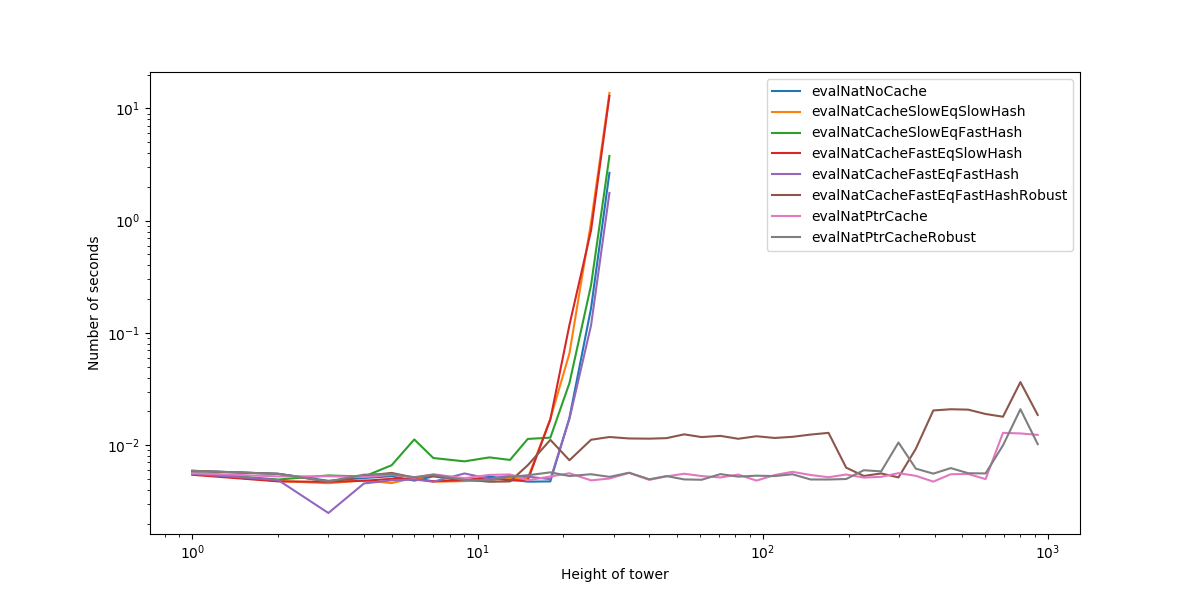}
  \caption{Comparing eight different variants of \lstinline{evalNat} on the tower of Figure~\ref{fig:twotowersdisjoint} that consists of two pointer-disjoint towers with a shared head.}
  \label{fig:experiments:twotowers}
\end{figure}

The first tower we consider is the simple maximally-shared tower of Figure~\ref{fig:tower}.
The results are shown in Figure~\ref{fig:experiments:tower}, with both axes in log-scale.
We see that as expected, the first four variants all take exponential time, whereas the latter four remain linear in \lstinline{n}.
The second tower we consider is that of Figure~\ref{fig:twotowersdisjoint}, which consists of two pointer-disjoint towers
with a shared head. The results are shown in Figure~\ref{fig:experiments:twotowers}, again with both axes in log-scale.
In contrast to Figure~\ref{fig:experiments:tower}, here we see that the \lstinline{evalNatCacheFastEqFastHash} variant of \S\ref{sec:traverse:perfect}
takes exponential time, for the reasons discussed in \S\ref{sec:traverse:perfect}. The last three variants still take time linear in \lstinline{n}, as expected.
Code to reproduce all experiments is included in the supplementary material.

\section{Discussion}\label{sec:discussion}

A reduced-order binary decision diagram (ROBDD) is a canonical example of a datastructure that
requires maintaining some kind of max-sharing invariant, \ie that if two nodes in a graph are structurally equal
then they must have the same unique identifier, where the identifier could be either an integer or a memory address.
We note that in contrast to the problems we have considered in this paper,
existing pure languages can construct ROBDDs from scratch and manipulate them without exponential blowup, \eg
by either of the two pure approaches used by~\cite{braibant2014implementing} to implement them in Gallina.
The important distinction is that in existing pure languages, one can easily build ROBDDs
\emph{from the bottom up} using an explicit graph representation,
whereas if you \emph{start} with a term whose tree size is astronomically large,
there is nothing you can do without the ability to compare memory addresses of subterms.
It is also common practice within compilers to build and maintain compact representations of programs,
e.g. with aggressive \lstinline{let}-abstractions. This bottom-up style is appealing when it applies,
but it is not feasible in interactive theorem provers. In contrast to compilation, where the
input programs for the compiler stack are generally written explicitly by humans rather than being the output of other (meta-)programs,
terms in interactive theorem provers are often the result of long chains of arbitrary, user-written meta-programs.
There is no way to circumvent the need to exploit sharing in term trees without severely limiting the convenience
or expressivity of the meta-programming frameworks.

In contrast to Lean which is directly compiled to C and which has its own runtime,
Gallina code is generally executed by first extracting it to OCaml
and then compiling the resulting OCaml program.
The standard way of augmenting Gallina programs with access to impure features
is to specify that particular Gallina functions should be extracted to particular
(possibly impure) OCaml functions.
This process is ad-hoc and unsafe in general, as the system itself cannot discern
pure extraction instructions from impure ones.
For example, \citet{braibant2014implementing} implement a naive BDD type in Gallina,
extract it to an OCaml type that stores a unique identifier,
extract the Gallina constructors to OCaml ``smart'' constructors
that make use of a hash-consing library to guarantee maximal sharing,
and extract the structural equality test on their BDD type to OCaml's physical (\ie pointer)
equality test. Thus when they execute their program,
equality between BDDs is determined by comparing pointers only.
However, their meta-logical soundness argument is subtle,
and requires that the regular OCaml constructors are never used directly.
Moreover, they give an example of a tempting smart constructor that would
introduce inconsistencies between the original Gallina and extracted OCaml code.
In contrast, the abstractions we have introduced can be used freely
by users without any risk of impurity.

Pointer equality is a particularly delicate issue in Haskell.
There are several reasons why an object may not even have the same address as itself,
for example
it might get duplicated during garbage collection,
or it may live in two different un-evaluated thunks.
In part because of these issues, checking pointer equality in Haskell is considered not only unsafe but ``really'' unsafe:
indeed, the operation is named \lstinline{reallyUnsafePtrEquality#}.\footnote{\url{https://downloads.haskell.org/~ghc/8.8.2/docs/html/libraries/ghc-prim-0.5.3/GHC-Prim.html}. Accessed 2/21/2020.}
To support an analogue of memory addresses with more desirable properties,
\citet{jones1999stretching} introduce the \emph{stable name} abstraction for Haskell that allows fast equality, comparison, and hashing,
and that is guaranteed to be stable over the lifetime of an object.
However, creating a stable name for an object is not a pure operation, since \eg the stable names of two objects might compare differently on different runs,
and so the creation of stable names is still forced to be the \lstinline{IO} monad.

\citet{goubault1994implementing} proposed a runtime system for a functional language that
would hash-cons all values to ensure maximal sharing at all times.
The language could then have built-in support for datastructures such as maps
that use memory addresses for ordering and equality.
However, despite the promising empirical results reported in the paper,
there is a general consensus that hash-consing is slow and wasteful on many workloads,
especially for functional programming where it is particularly common to produce many transient objects.
We also remark that several functional programming languages including Lean4,
PVS~\cite{owre1992pvs}, SISAL~\cite{sisal}, and SAC~\cite{sac}
have support for transforming functional array updates into destructive ones using reference counts,
and hash-consing arrays would introduce undesired sharing and so prevent destructive updates from being applied.
Hash-consing arrays is also inefficient in general, since the cost is linear in the size of the array.

Lastly, the ACL2 theorem prover~\cite{kaufmann1997industrial} is based on a subset of Common Lisp~\cite{steele1990common}
and includes structural equality as a primitive. Thus the runtime may choose to (recursively) accelerate it by short-circuiting in the pointer-equal case.
\citet{boyer2006function} also introduced a hash-consing framework for ACL2 akin to an opt-in version of that proposed by \cite{goubault1994implementing},
with a new primitive \lstinline{hons} that is a hash-consing version of the standard \lstinline{cons} and a new primitive \lstinline{hons-equal}
that is like \lstinline{equal} but that may use the hash-cons table to accelerate the check.
Whereas ACL2 builds a fixed set of pointer-based optimizations into the language,
our approach allows users to safely implement their own pointer-based optimizations on their own datatypes.

\section{Conclusion}\label{sec:conclusion}

We have presented a new way to use dependent types to seal many pointer-based optimizations
behind pure functional interfaces while requiring only a negligible amount of additional trust.
We introduced primitives for conducting pointer equality tests (\lstinline{withPtrEq} and \lstinline{withPtrEqResult}),
for sharing the common data across terms of arbitrary types (\lstinline{withShareCommon}), and for
directly observing pointer addresses (\lstinline{withPtrAddr}).
In all cases, the low-level imperative implementations of these primitives
are functionally indistinguishable from their pure reference implementations.
We also showed how to use these new primitives to achieve exponential speedups when traversing heavily-shared
terms. We believe our work constitutes a significant step towards making pure functional programming
a viable option for building high-performance systems for automated reasoning.

\begin{acks}
  We thank Sebastian Ullrich and Ryan Krueger for helpful feedback on early drafts.
\end{acks}

\bibliography{pureptr}

\end{document}